\documentclass[11pt]{article}
\usepackage{setspace}
\usepackage{amssymb}
\usepackage{amsmath}
\usepackage{amsfonts}
\usepackage{slashed}
\usepackage[symbol]{footmisc}

\def\be{\begin{equation}}

\def\ee{\end{equation}}

\def\dd{\partial}

\def\bea{\begin{eqnarray}}

\def\eea{\end{eqnarray}}

\newcommand\eps{\epsilon}

\makeatletter
\def\blfootnote{\xdef\@thefnmark{}\@footnotetext}
\makeatother

\begin{document}

\singlespace

\vspace*{.3in}

\begin{center}

{\Large\bf Renormalizability Properties of Supergravity\footnote{Supported in part by National Science Foundation Grant No. PHY-76-07299.}}

{\large S.\ Deser,} {\large J.\ H.\ Kay\footnote{National Science Foundation Graduate Fellow},} {\large K.\ S.\ Stelle}

{\it Physics Department,  Brandeis University, Waltham, MA 02154 \\
}

\end{center}

{\it \noindent This pre-arXiv work, reproduced from the original, PRL 38 (1977) 527 (submitted 27 Dec. 1976), introduced the now ubiquitous $3$-loop ``$R^4$" supersymmetric counterterm, key to the UV finiteness issue in all supergravity theories, particularly the much discussed $N=8$ case. It also established the basic pattern for the corresponding $4$-point finite corrections in string theory effective actions. We thank J Franklin for producing this arXiv version.
}

\begin{abstract}
The possible local counterterms in supergravity are investigated to all loop orders. Supersymmetry implies that (1) supergravity-matter coupling is one-loop nonrenormalizable, with a specific counterterm; (2) pure supergravity is renormalizable at both one and two loops; (3) it fails at three loops; (4) extended supergravity models may avoid the three-loop catastrophe, and have no dangerous local counterterms to any order. In that case, the nonleading divergences could be removed by field redefinitions, which would establish renormalizability for these systems.
\end{abstract}

One of the chief motivations in the construction of supergravity [1,2] was the hope that this theory, in contrast to ordinary Einstein gravity [3],  would
prove renormalizable. The gravity-spin-2 system should be better behaved by virtue of the additional constraints imposed by local 
supersymmetry on the possible counterterms. (The existence of Gauss-Bonnet-like identities, which will be given below, in the spinor sector of supergravity was an early indication [2] of this possibil- ity. )

We shall show here that this hope becomes progressively better founded as one ranges over different versions of supergravity. Pure supergravity is better behaved than pure gravity. The latter is only one one-loop renormalizable, while the former is completely good through two loops. However, from the three-loop level onward, there appear possible superinvariant local 
counterterms which do not vanish on the mass shell.  These terms are formally related to those, which we shall also exhibit, already occurring at the one-loop level for supergravity coupled to matter multiplets such as (1, 2), but may, in turn, be avoidable in extended supergravity models [4] through their additional internal symmetries.

 Throughout, we assume that any meaningful regularization procedure always preserves, without anomaly, gauge invariance of the second kind,
 so that our analysis of supersymmetry invariants is effectively that of all possible counterterms.
Before proceeding, we summarize what is already known concerning renormalizability. Explicit calculation has established that supergravity coupled to the $(1, 1/2)$ multiplet is one-loop non-
renormalizable in the four-photon sector [5].  By contrast, in pure super gravity, at both one-loop [6] and two-loop [7] levels, and in $O(2)$ extended supergravity at one-loop [6], there are no problems on shell through four-particle amplitudes. These results were obtained by S-matrix arguments involving helicity conservation in presence of global supersymmetry.  
 
\subsection*{Notation and basic lemmas}
The supergravity action has the first-order form [2]
 \begin{equation}
 \begin{aligned}
 I &= \frac{1}{2} \int (d^4 x) \, \left[ {\mathcal R}(e,\omega) + i \, \bar \psi_\mu \, \gamma_5 \, \gamma_\nu \ ^* f^{\mu\nu}\right], \\
 \ ^* f^{\mu\nu} &\equiv \frac{1}{2} \, \eps^{\mu\nu\alpha\beta} \, f_{\alpha \beta}, \, \, \, \, \, \, \, \, \, \, \, f_{\alpha\beta} \equiv D_\alpha \psi_\beta - D_\beta \psi_\alpha, \\
 D_\alpha &\equiv \dd_\alpha - \frac{1}{2} \, \omega_{\mu\alpha\beta} \, \sigma^{\alpha\beta}, \, \, \, \, \, \, \, \, \, \, \, [D_\alpha,D_\beta] = -\frac{1}{2} R_{\alpha\beta}^{\, \, \, \, \, } \cdot \sigma.
 \end{aligned}
 \end{equation}
The field equations defining the common mass shell can be written in various useful ways.  For the spinor field, they have the formally duality-invariant [8] structure
\begin{equation}
\gamma_\mu \ ^*f^{\mu\nu} = \gamma^\mu \, f_{\mu\nu} + \gamma_5 \ ^* f_{\mu\nu} = 0.
\end{equation}
The fact that $f$ may be replaced by $-\gamma_5 \ ^* f$ on shell [like $D_{\mu\nu} \lambda \equiv (\gamma_\mu \dd_\nu - \gamma_\nu \dd_\mu) \, \lambda = \frac{1}{2} \, \gamma_5 \, \eps_{\mu\nu}^{\, \, \, \, \alpha\beta} \, D_{\alpha\beta} \lambda$ for spin $\frac{1}{2}$] is very useful, as are the free-field relations 
$\dd_\mu f^{\mu\nu} = \dd_\mu \ ^*f^{\mu\nu} = \slashed\dd f^{\mu\nu} = 0$.    The stress tensor $T^{\mu a}(\psi)$ is traceless on shell, so that also $R=0$. 
For the free field, the following relations hold on shell:
For any bilinear form, $\bar f_{\mu\nu} \, \gamma_\lambda \dd^{(m)} f_{\alpha\beta} = -\bar f_{\alpha\beta} \, \gamma_\lambda \, \dd^{(m)} f_{\mu\nu} + X$, where $X$ involves at least one contraction between $\bar f$ and $f$ indices. It follows further that any $\bar f \, \gamma \, \dd^{(m)} \, f$ terms are ultimately reducible to contracted ones, which have the property that
$\bar f_{\lambda \mu} \, \gamma_\alpha \, \dd^{(m)} f_\nu^{\, \, \lambda}$ is symmetric in $(\mu\alpha\nu)$.

Since our arguments will also make use of the global supersymmetry of the free $(2, 3/2)$ multiplet,
we record the corresponding (weak-field) trans- formations on (free) shell:
\begin{equation}
\begin{aligned}
\delta \omega_{\mu a b} &= - i \bar{\alpha} \gamma_\mu f_{ab}, \, \, \, \, \, \, \, \, \, \, \, \delta \psi_\mu = -\omega_\mu \cdot \sigma \alpha, \\
\delta f_{\mu\nu} &= -R_{\mu\nu} \cdot \sigma \alpha, \, \, \, \, \, \, \, \, \, \, \, \delta R_{\mu\nu\alpha\beta} = i \, \bar{\alpha}\, D_{\mu\nu} \, f_{\alpha\beta},
\end{aligned}
\end{equation}
where $\alpha$ is the (anticommuting) transformation parameter. We shall use the obvious but invaluable fact that since supergravity is supersymmetric, its field equations transform covariantly into each other, so that variations of quantities which vanish on shell also vanish there. We shall exploit this fact to do all of our work on shell (where everything simplifies enormously) since we are only interested in possible nonvanishing on-shell invariants. Although those parts of an invariant which vanish on shell do so by virtue of the full nonlinear field equations, any surviving terms must achieve invariance on shell by starting again from linearized supersymmetry and building up. In this process, the field equations may now, of course, be used, but to lowest order only their linearized parts are relevant. As explained below, this means that it will be sufficient to consider only leading terms which are manifestly invariant under the transformation
$\delta \psi_\mu = 2 \, \dd_\mu \alpha$, and so involve only $f_{\mu\nu}$ rather than $\psi_\mu$, and correspondingly for curvature versus affinity.

Finally, we note the ``super" Gauss-Bonnet theorem which extends the usual relation $I_1 = I_2 + I_3$, where
\begin{equation}
I_1 = \int R_{\mu\nu\alpha\beta}^2, \, \, \, \, \, I_2 = \int (4 \, R_{\mu\nu}^2 - 2 \, R^2), \, \, \, \, \, I_3 = \int R^2,
\end{equation}
to the global symmetry partners
\begin{equation}
K_1 = -2 \, i \, \int \bar f^{\alpha\beta} \slashed \dd f_{\alpha\beta}, \, \, \, \, \, K_2 = 4 \, i \int \bar f^{\mu\nu} \gamma_\nu \slashed  \dd \gamma^\beta \, f_{\beta\mu}, \, \, \, \, \, 
K_3 = -4 \, i \, \int \bar f \cdot \sigma \slashed \dd \sigma \cdot f,
\end{equation}
according to $K_1 = K_2 + K_3$ at lowest order.  Note also that the above relations are unaltered for the parts quadratic in the fields, if arbitrary powers of the D'Alembertian are inserted between the factors in $I_i$ and $K_i$.

\subsection*{One loop}
The Gauss-Bonnet results tell us that there are just two globally symmetric arrays of scale dimension four, namely $I_2 + K_2$ and $I_3 + K_3$.  The Noether prescription augments these with terms of the form $\bar f \gamma \psi R$, $\bar \psi \gamma \psi \dd\ R$.  We merely state the result: They build up the original $I_i + K_i$ into their full on-shell form, e.g., $\sim \bar f \cdot \gamma \slashed D \gamma \cdot f + [R_{\mu\nu} - T_{\mu\nu} ]^2 + \bar f \cdot \gamma \psi R_{\mu\nu}$ through $\psi^2$ terms.  Note that since we are interested in the values of all invariants rather than their variational properties, it is gratifying that the torsion contributions
build up correctly, and our terms are all bilinear in $\delta I/\delta e_{\mu a}$ and $\delta I/\delta \psi_\mu$.  Rather than attempt this lengthy procedure to the bitter $(\psi^8)$ end, we now present the general argument which shows that there are no other invariants on shell. Any surviving terms beyond those needed to complete the above on-shell (vanishing) arrays would them- selves have to be, in their lowest-order parts, invariant on shell under global supersymmetry
and the separate spin-$2$ and -$3/2$ Abelian gauges. But the purely gravitational terms are already included in our two arrays, as are the quadratic purely fermionic terms (since the $K_i$ are the only independent ones). As for three-point terms, we emphasize that since the two-point starting expressions all vanish as bilinears in the field equations, the three-point terms generated via Noether coupling automatically vanish on shell.
It is only in such three-point terms that one could conceivably have an Abelian invariance which was not manifest, but required use of the linearized
field equations. Such terms would necessarily be of the form $\int J \psi$, with $J^\mu$ a conserved current of the linearized theory and thus a bilinear structure (compare $\int F \cdot A \times A$ in Yang-Mills theory).
All other three-point terms would have to be manifestly Abelian invariant, but there are insufficient derivatives to ensure this. Similarly, the four- and higher-point terms, schematically
$\psi^4 \dd^2$, $\psi^6 \dd$, $\psi^8$, would have to be manifestly Abelian and global invariant, and again there are too few derivatives. Therefore the one-loop divergences (which are, of course, entirely of leading local type) all vanish on shell for arbitrary numbers of particles.

\subsection*{Two loops}
Our basic starting point here is
the same set of two arrays which begin quadratically in gravitons, namely $I_i +K_i$ with a D'Alembertian insertion.  All other initially quadratic invariants $\int R \ldots DD R \dots$ 
reduce to these plus terms $\sim \int R \ldots ^3$.  As in one loop, the Noether coupling produces three-point terms which vanish on shell,
There are now, however, sufficient derivatives to construct three-point Abelian invariants like
$R \ldots$, $R \bar f \, \gamma \, \dd f$. While not part of the Noether coupling, such terms could in principle be needed to restore global supersymmetry to this order.
However, when we go on shell, the surviving parts of these terms would have to be invariant.
But the $R \ldots ^3$ term has a variation which requires a companion of the form $R^{\mu\nu\alpha\beta} \left(\bar f^\lambda_{\, \, \mu} \, \gamma_\alpha \, \dd_\beta \, f_{\lambda\nu}\right)$,
while, as we have seen, the $\bar f_\gamma \, \dd f$ is (on shell) necessarily symmetric in $\mu\alpha\nu$ so that it annihilates the $R_{\mu\nu\alpha\beta}$ by index symmetries, and either
partner with contracted indices vanishes on shell.  Finally, there can be no terms $\sim (\bar f \, f)(\bar f\, f)$ because
they have no possible coordinate-invariant partners,  purely by derivative power counting. To summarize, there are no nonvanishing on-shell
two-loop invariants because no appropriate terms, aside from the vanishing arrays, can be constructed
as global starting points.

\subsection*{Gravity supermatter}
We indicate here how our methods imply that the coupled supergravity-matter $(1,1/2)$ multiplet [9] is one-loop divergent;
this is actually a foreshadowing of the three-loop problems below. At the global level, any multiplet obeys [10] the following on-shell transformations
(up to divergences) of its total stress $t_{\mu\nu}$, supercurrent $j_\mu$,  and axial current $c_\mu$ (with our conventions):
\begin{equation}
\begin{aligned}
\int t^{\mu\nu} \, \delta t_{\mu\nu} &= i \, \int t^{\mu\nu} \, \bar \alpha \, \gamma_\mu \slashed\dd j_\nu, \, \, \, \, \, \, \, \, \, \, \, \delta c_\mu = i \bar \alpha \, \gamma_5 \, j_\mu, \\
\delta j_\mu &= t_{\mu\nu} \, \gamma^\nu \, \alpha + \gamma_5 \, \slashed \dd \, c_\mu \, \alpha - \frac{1}{2} \ ^*D_{\mu\nu} \, c^\nu \alpha.
\end{aligned}
\end{equation}
Thus, on shell, the quantity
\begin{equation}
\Delta_1 I = \int \left[ t_{\mu\nu}^2 + i \, \bar j^\mu \slashed \dd j_\mu - \frac{3}{2} \, c^\mu \Box c_\mu \right]
\end{equation}
is globally invariant. For supergravity itself, it is excluded because the components [e.g., the
Einstein pseudotensor or $T_{\mu\nu}(\psi)$] lack local Lorentz and Abelian invariance, but it is permitted
for the $(1,1/2)$ system. Since the $(2,3/2)$ and $(1,1/2)$ are globally independent, (7) is perfectly allowed as a starting point [11] despite the absence (on
shell) of a $(2,3/2)$ partner. Indeed, one can show, using also Maxwell duality invariance [8], that this
counterterm determines all the possible one-loop
divergences; these include in particular the four-photon
amplitude $\sim (T_{\mu\nu}^{\, \, \, \, \, M})^2$ from the Maxwell
stress tensor which has been found by explicit calculation [5].  This problem is clearly traceable
to having two separate groups, and we shall see how it is avoidable in extended supergravity, i.e.,
why (7) is forbidden there even for the lower spin parts of an $O(N)$, also in agreement with the $O(2)$
calculations [6].

\subsection*{$n$-loop supergravity}   
The generic local $n$-loop
counter terms may again be treated by separating
the on-shell vanishing arrays from higher particle terms. We only wish to indicate here the
basic requirement for nonvanishing $n$-loop invariants to exist by considering, say, $R^{n+1}$ terms.
Their variations are of the form $[\sum(R^n)]_{\mu\nu\alpha\beta} \times \bar \alpha D^{\mu\nu} \, f^{\alpha\beta}$. Their partners would have to have
the form $R^{n-1} \bar f \gamma \dd f$.  But variation of the latter would always yield extra terms involving $\dd(R^{n-1}) \times R \bar \alpha \gamma f$,
which would have to vanish separately because they could not be cancelled by any other
covariant term (explicit affinities would be required). (Identically conserved quantities run
into other difficulties.) Further, terms like $R^n \dd^2$ cannot work either, because one could iteratively reduce these to terms with similar difficulties.
Likewise, any higher $R^m (\bar f f)^n$ terms would eventually couple to the above types. Thus it would appear that renormalizability hinges on the absence
of conserved geometrical tensors at least
quadratic in curvature. Unfortunately, these do
exist, as we now discuss, and first affect three
loops.

\subsection*{Three-loop supergravity}
The Bel-Robinson tensor [12] is defined by
\begin{equation} 
T_{\mu\nu\alpha\beta} = -[ R^{\lambda\, \, \, \rho}_{\, \, \, \alpha \, \, \, \mu} \, R_{\lambda\beta\rho\nu} + \ ^*R^{\lambda\, \, \, \rho}_{\, \, \, \alpha \, \, \, \mu} \, \ ^*R_{\lambda\beta\rho\nu}].
\end{equation}
It is totally symmetric, traceless, and conserved on shell. We now assert that the following quantity
is on-shell invariant, but nonvanishing:
\begin{equation}
\Delta_2 I = \int \left\{ [T_{\mu\nu\alpha\beta} + H_{\mu\nu\alpha\beta}]^2 + i \, \bar J^{\mu\alpha\beta} \slashed\dd  J_{\mu\alpha\beta} - \frac{3}{2} \, C^{\mu\alpha\beta} \, \Box C_{\mu\alpha\beta} \right\},
\end{equation}
where
\begin{equation}
H_{\mu\nu\alpha\beta} = -(i/2) \bar f^\lambda_{\, \, \alpha} \, (\gamma_\mu \dd_\nu + \gamma_\nu \dd_\mu) \, f_{\beta\lambda}, \, \, \, \, \, \, \, \, \, \, \,
J_{\mu\alpha\beta} = R^\lambda_{\, \, \alpha} \cdot \sigma \gamma_\mu f_{\lambda\beta}, \, \, \, \, \, \, \, \, \, \, \,
C_{\mu\alpha\beta} = -(i/2) \bar f^\lambda_{\, \, \alpha} \gamma_5 \gamma_\mu f_{\beta\lambda}.
\end{equation}
This is because essentially the same rules hold for these objects as for the $t^2$ multiplet of (6):
\begin{equation}
\begin{aligned}
\int(T + H)^{\mu\nu\alpha\beta} \delta(T + H)_{\mu\nu\alpha\beta} &= i \, \int (T + H)^{\mu\nu\alpha\beta} \, \bar \alpha \, \gamma_\mu \slashed \dd J_{\nu\alpha\beta}, \\
\delta C_{\mu\alpha\beta} = i \, \bar \alpha \gamma_5 \, J_{\mu\alpha\beta}, \, \, \, \, \, \, \, \, \, \, \,
\delta J_{\mu\alpha\beta} &= ([T+H]_{\mu\nu\alpha\beta} \gamma^\nu) \alpha + (\gamma_5 \slashed \dd C_{\mu\alpha\beta}) \, \alpha - (\frac{1}{2} \ ^*D_{\mu\nu} \, C_{\alpha\beta}^{\, \, \, \, \, \, \nu}) \, \alpha.
\end{aligned}
\end{equation}
Of course, we do not know whether (9) is part of a locally supersymmetric off-shell invariant, nor
whether some miraculous cancellation might kill it in explicit calculations, but these escapes seem
highly unlikely in view of the close parallel to one-loop supermatter. Furthermore, invariants
analogous to $\Delta_2 I$ may be constructed in higher loops using more derivatives through the Zilch procedure [13].
This class of invariants seems to be exhaustive: There are no dynamically conserved
tensors cubic or higher in the curvature (i.e., no cubic conserved currents for a free
field).

\subsection*{Extended supergravity} 
Very recently, it has been shown [4] that extensions of pure supergravity into a larger, rigid, single, global multiplet exist,
with internal symmetries such as $O(N)$, $N \le 8$. As stated above, these models avoid the one-loop disaster of supermatter coupling, because
one cannot use the form (7) constructed from, e.g., their lowest multiplet part without violating the rotation invariance, and we believe
the same argument will exclude the corresponding three-loop disaster. The point (which may even hold in presence of extended supermatter)
is that the transformation rules, Eqs. (6) and (10), do not commute with the internal rotations
and therefore there is no analog of (9). Should this be the case on detailed examination of specific
$O(N)$ models, they would have no locally constructed invariants [14] which do not vanish on shell,
to any loop order.

\subsection*{\it Field redefinitions} 
lt was already noted by 't Hooft and Veltman [3] that the one-loop divergences $\eps^{-1} [a R_{\mu\nu}^2 + b \, R^2]$ in pure gravity could be
absorbed by an unconventional renormalization of the metric $\bar g_{\mu\nu} \sim g_{\mu\nu} + \eps^{-1} (R_{\mu\nu} + g_{\mu\nu} \, R)$ since $\int R(\bar g) \sim
\int R(g) + \int (\delta R/\delta g) \, \delta g$.  This procedure can clearly be followed order by order to absorb any local invariants which vanish on shell, by redefining
both the gravitational and $\psi$ fields appropriately.
This somewhat unorthodox renormalization will systematically shrink away divergent subintegrals, leaving only the local divergences, which
we have already dealt with [15].  There seems to be no reason to doubt, at least in perturbation theory, that the $S$ matrix is unchanged by such field
redefinition, but this point clearly deserves careful study.

%\subsection*{Acknowledgements}
%SD was supported in part by Grants NSF PHY- 1266107 and DOE \# DE-SC0011632. 

\end{document}